\documentclass{aa}
\usepackage{graphicx}
\usepackage{txfonts}
\usepackage{natbib}
\begin{document}
   \title{In search of an unbiased temperature estimator \\ for statistically poor X-ray spectra}

   \author{Alberto Leccardi
          \inst{1,2}
          \and
          Silvano Molendi
          \inst{2}
          }


\institute{
          Universit\`a degli Studi di Milano, Dip. di Fisica, via Celoria 16,
          I-20133 Milano, Italy
          \and
          INAF-IASF Milano, via Bassini 15,
          I-20133 Milano, Italy
          }


   \abstract{
      Although commonly employed by X-ray astronomers, maximum likelihood estimators are known to be biased.
      In this paper we investigate the bias associated to the measure of the temperature from an X-ray
      thermal spectrum.
      We show that, in the case of low surface brightness regions, commonly adopted estimators, such as
      those based on $\chi^2$ and Cash statistics, return strongly biased results.
      We stress that this can have strong implications when measuring the temperature of cluster outer
      regions with current experiments.
      We consider various approaches to overcome this problem, the most effective is a technique which
      allows us to correct the bias a posteriori.
      Extensive montecarlo simulations show that our correction returns excellent results under different
      conditions.
   \keywords{Methods: statistical -- X-rays: galaxies: clusters }
   }

   \maketitle
%

\section{Introduction}

With the advent of \emph{XMM-Newton} and \emph{Chandra} it has become possible to explore to a
certain extent the physical properties of the intra-cluster medium (ICM) in the outer
regions of galaxy clusters.
In dealing with these regions, there are both statistical and systematic issues which need to be
addressed: typically, the spectra have poor statistic (i.e. few counts/bin) and a high
background, especially at high energies, where, also due to the sharp decrease of the
effective area of the experiments, the instrumental background dominates over other components.
In this paper we employ simulations to examine how best to analyze this kind of spectra.
Here we focus on the treatment of statistical errors and neglect systematic ones, which will
be discussed in a forthcoming paper \citep{leccardi07}.
More specifically the question we wish to address is the following: \emph{``What are the effects
of pure statistical uncertainties in determining interesting parameters of highly non
linear models (e.g. the temperature of the ICM), when we analyze spectra accumulated from low
surface brightness regions using current X-ray experiments?''}
To address this question, we perform a set of simulations: first, we choose the input values for
model parameters and produce the expected spectrum; then, we generate a large number of perturbed
spectra representing a large set of measurements; finally, we analyze them with different techniques
based on the method of maximum-likelihood (hereafter ML) and compare the results.
Our choices of simulation parameters (e.g. spectral model, energy band, fixed parameters, etc.)
are justified by our practical issue, i.e. the determination of the temperature in the outer
regions of massive galaxy clusters.
Our analysis is mainly focused on \emph{XMM-Newton}, however most of our results are valid
in all cases when analyzing low count Poisson-distributed data.

From a more general perspective, ours may be viewed as an attempt to quantify the significance
of the bias of ML estimators commonly adopted by X-ray astronomers to determine spectral parameters.
As we shall see, the most common ML estimators, indeed all those available within XSPEC\footnote{We used
XSPEC 11.3 (http://heasarc.nasa.gov/docs/xanadu/\\xspec/xspec11/index.html)}, are characterized by
a substantial bias when applied to our specific case.
A long term solution to the problem requires an unbiased, or perhaps a less biased, estimator
to be found and implemented within standard fitting packages (i.e. XSPEC).
Another, faster, solution involves correcting the bias a posteriori making use of extensive
montecarlo simulations.

The outline of the paper is the following.
In $\S$\ref{sec: ideal} we consider the idealized source only case.
In $\S$\ref{sec: real} we include the background considering two cases: in the first the source
contribution is much more important than the background one, in the second the opposite is true.
In $\S$\ref{sec: correct bias} we try to correct the bias: in $\S$\ref{subsec: estimators} we search
for a long term solution of the problem using different estimators; in $\S$\ref{subsec: lognorm} and
$\S$\ref{subsec: triplet} we correct the bias a posteriori, in particular in
$\S$\ref{subsec: triplet} we explain how to use our ad-hoc recipe
(i.e. the so-called ``triplet'' method).
In $\S$\ref{sec: conclusions} we summarize our main results.

Errors are quoted at one sigma for one interesting parameter, unless otherwise stated.


\section{The source only case}  \label{sec: ideal}

In this section we deal with the idealized source only case.
We represent the source with an absorbed thermal model (\verb|WABS*MEKAL| in XSPEC).
The parameter values are the following: the equivalent hydrogen column density along the
line of sight, $N\mathrm{_H}$, is $2.5\times10^{20}$~cm$^{-2}$; the metallicity, $Z$, and
the redshift, $z$, are respectively 0.25 solar and 0.2; the temperature, $kT$, is
7~keV and the normalization, $N\mathrm{_S}$, is $3.5\times10^{-3}$ in XSPEC units.
Our redistribution matrix (RMF) and effective area (ARF) have been produced from the
observation number 0093030101 of the galaxy cluster Abell~1689 with the EPIC-MOS1
instrument; the angular size of the region for which we accumulate the simulated spectra
is about 4 arcmin$^2$, which corresponds to the ring between 1.0\arcmin~and
1.5\arcmin~centered on the cluster emission peak.
The exposure times considered are 5~ks, 10~ks, 100~ks and 1\,000~ks and the total counts in the
2.0-10.0~keV band are respectively about 600, 1\,200, 12\,000 and 120\,000.
For each channel we perturb the number of counts with a Poisson distribution centered on the
expected value. We repeat this step for $N\mathrm{_{meas}}$ times (with $N\mathrm{_{meas}}$ very
large) to obtain $N\mathrm{_{meas}}$ spectra, which simulate $N\mathrm{_{meas}}$ independent
measurements of the source.

We fit simulated spectra using the $\chi^2$ and the Cash statistics, where the latter
is more suitable to analyze spectra with few counts per channel \citep{cash79,nousek89,mighell99,arzner06}.
We recall that each measurement can be represented by the number of counts, $O\mathrm{_i}$,
observed in each channel $i$ ($i=1,...,N$ where $N$ is the number of channels).
The probability, $Q$, of obtaining this particular measurement (i.e. this particular spectrum) is the
product of Poisson distributions and can be expressed as a function of the expected counts, $E\mathrm{_i}$,
which depend\footnote{In the following equations the dependency of $E\mathrm{_i}$ from $\alpha$ is omitted
for clarity.} on the particular set of model parameters, $\alpha$ (e.g. in this case
$\alpha=(N\mathrm{_H},kT,Z,z,N\mathrm{_{S}})$):
\begin{equation}
Q(\alpha) = \prod_{i=1}^N \frac{E\mathrm{_i}^{O\mathrm{_i}} \: \mathrm{e}^{-E\mathrm{_i}}}{O\mathrm{_i}!} \; .
\label{eq: poiss}
\end{equation}
The associated log-ML function $C$ \citep{cash79} is defined as follows:
\begin{equation}
C(\alpha) = -2 \; \ln Q(\alpha) = -2 \; \sum_{i=1}^N \; \left( O\mathrm{_i} \: \ln E\mathrm{_i} - E\mathrm{_i} - \ln O\mathrm{_i}! \right) \; .
\label{eq: C}
\end{equation}
The best set of parameters is determined by maximizing $Q$ (i.e. minimizing $C$) with respect to $\alpha$.
Conversely, the $\chi^2$ statistic is based on the hypothesis that each spectral bin contains a sufficient
number of counts that the deviations of the $O\mathrm{_i}$ from the $E\mathrm{_i}$ have a Gaussian
distribution.
This hypothesis is satisfied for large $O\mathrm{_i}$, when $Q$ can be approximated by a product of
Gaussian distributions and the associated log-ML function $\chi^2$ is defined as follows:
\begin{equation}
\chi^2(\alpha) = \sum_{i=1}^N \; \frac{ \left( O\mathrm{_i} - E\mathrm{_i} \right)^2}{\sigma\mathrm{_i}^2} \; ;
\label{eq: chi}
\end{equation}
where $\sigma\mathrm{_i}$ is usually the uncertainty in the i-th bin
($\sigma\mathrm{_i}$~=~$O\mathrm{_i}^{1/2}$).
The larger $O\mathrm{_i}$, the better the approximation of Gaussian regime.
Channel grouping is a widely used strategy that allows to reduce the bias introduced by this approximation.
We group channels in order to have at least 25 counts per bin, which is a commonly adopted compromise.
Conversely, when using the Cash statistic we perform a minimal grouping to avoid channels with
no counts, i.e. the spectrum is substantially unbinned and no spectral information is lost.
Each spectrum is fitted between 2.0 and 10.0~keV (the energy band we are interested in) with the
absorbed thermal model mentioned above.
The $N\mathrm{_H}$ is fixed to the input value (typical values of $N\mathrm{_H}$
for cluster observations have negligible effects above 2~keV), $z$ is allowed to vary
between 0.186 and 0.214 ($\pm7\%$ of the input value), while $kT$, $Z$ and $N\mathrm{_{S}}$ are free.
We determine best fit values and one sigma uncertainties for all parameters.

\begin{table}
  \caption{Weighted averages of temperature best fit values compared to the input value and relative
  differences $\Delta T/T_0$, using different exposure times and statistics.}
  \label{tab: ideal1}
  \centering
  \begin{tabular}{r l c r c r}
    \hline \hline
     & & \multicolumn{2}{c}{$\chi^2$} & \multicolumn{2}{c}{Cash} \\
    Exp. $\mathrm{^a}$ & $kT_0$ $\mathrm{^b}$ & $kT$ $\mathrm{^c}$ & $\Delta T/T_0$
    $\mathrm{^d}$ & $kT$ $\mathrm{^c}$ & $\Delta T/T_0$ $\mathrm{^d}$ \\
    \hline
    1000\, & 7.00 & 6.89$\pm$0.01 & -1.6\% & 7.00$\pm$0.01 & +0.0\% \\
    100\,  & 7.00 & 6.83$\pm$0.01 & -2.4\% & 7.03$\pm$0.01 & +0.4\% \\
    10\,   & 7.00 & 6.76$\pm$0.03 & -3.4\% & 6.91$\pm$0.02 & -1.3\% \\
    5\,    & 7.00 & 6.59$\pm$0.04 & -5.9\% & 6.81$\pm$0.03 & -2.7\% \\
    \hline
  \end{tabular}
  \begin{list}{}{}
    \item[Notes:] $\mathrm{^a}$ exposure time in kiloseconds; $\mathrm{^b}$ input temperature
    in keV; $\mathrm{^c}$~measured temperature in keV; $\mathrm{^d}$ relative difference.
  \end{list}
\end{table}

In Table~\ref{tab: ideal1} we compare the weighted average of the $N\mathrm{_{meas}}$ measured
temperatures to the input value, $kT_0$~=~7~keV, for different exposure times and statistics.
$N\mathrm{_{meas}}$ is chosen in order to have similar uncertainties on the average
($N\mathrm{_{meas}}$~=~1\,200 for 5 and 10~ks, $N\mathrm{_{meas}}$~=~300 for 100 and 1\,000~ks).
In almost all cases, the true temperature is underestimated by a few percent and the effect
becomes more evident for shorter exposure times. 
We recall that both $\chi^2$ and Cash statistics are based on the ML method.
Although X-ray astronomers make extensive use of ML estimators, it is well known from the
literature (e.g. \citealp{cowan98}) that: 1) ML estimators may be biased, i.e. the expectation
value may be different from the true value of the quantity to estimate; 2) ML estimators are
usually gaussian and unbiased only in the asymptotic limit.
In the case at hand, the asymptotic limit is approached when the total number of counts
becomes large.
The results reported in Table~\ref{tab: ideal1} show that: 1) both ML estimators are biased;
2) both estimators are asymptotically unbiased; 3) the Cash estimator tends to the true value
more quickly than the $\chi^2$ one.

As we have just pointed out, the $\chi^2$ is significantly more biased than the Cash estimator
(i.e. the difference between the expected and the true value is greater).
This is because the approximation of gaussian regime fails for few counts per bin.
The obvious implication is that to improve the precision of the $\chi^2$ estimates we need to increase
the number of counts in each bin, $N\mathrm{_{bin}}$.
\begin{table}
  \caption{Weighted averages of temperature best fit values compared to the input value and relative
  differences $\Delta T/T_0$, using different channel groupings.}
  \label{tab: ideal2}
  \centering
  \begin{tabular}{r l c r}
    \hline \hline
    $N\mathrm{_{bin}}$ $\mathrm{^a}$ & $kT_0$ $\mathrm{^b}$ & $kT$ $\mathrm{^c}$ &
    $\Delta T/T_0$ $\mathrm{^d}$ \\
    \hline
    400 & 7.00 & 6.99$\pm$0.01 & -0.1\% \\
    100 & 7.00 & 6.95$\pm$0.01 & -0.7\% \\
    25  & 7.00 & 6.89$\pm$0.01 & -1.6\% \\
    \hline
  \end{tabular}
  \begin{list}{}{}
    \item[Notes:] $\mathrm{^a}$ counts per bin; $\mathrm{^b}$ input temperature in keV;
    $\mathrm{^c}$ measured temperature in keV; $\mathrm{^d}$ relative difference.
  \end{list}
\end{table}
In Table~\ref{tab: ideal2} we compare the results obtained using the $\chi^2$ with different channel
groupings (note that using the Cash statistic this is not necessary).
The input temperature is 7~keV, the exposure time is 1\,000~ks and the number of measurements is 300.
As expected, we find that the greater is the number of counts in each bin, $N\mathrm{_{bin}}$, the
smaller is the bias.
However, in practice, grouping of a large number of channels is not desirable, because it causes loss
of spectral information; 25 counts per bin is a commonly adopted compromise.
We have to mention the existence of an alternative way to reduce the bias which affects the $\chi^2$
estimator for few counts per bin.
Some authors (e.g. \citealp{churazov96,gehrels86,primini95}) choose different
statistic weights ($\sigma\mathrm{_{i}}$ in Eq.~\ref{eq: chi}) instead of the standard $O\mathrm{_{i}}^{1/2}$.
We re-analyzed our spectra using all the alternative weights implemented in XSPEC and we obtain results
rather similar to those already discussed for the Cash statistic.

In this section we have analyzed the ideal case of a thermal source without a background.
The results are summarized as follows.
The standard $\chi^2$ statistic works well only in the Gaussian regime, which is reached when performing
a strong channel grouping (see Table~\ref{tab: ideal2}).
When using a realistic grouping (e.g. 25 counts per bin) the measured temperature, $kT$, is lower than
the true temperature, $kT_0$ (see Table~\ref{tab: ideal1}).
The Cash statistic \citep{cash79} works better, because it is based on the ML function for Poisson
processes; however, when the spectra total number of counts is small, $kT$ is lower than $kT_0$ by a
few percent (see Table~\ref{tab: ideal1}).
This means that the Cash ML estimator is only asymptotically unbiased (for a review about
parameter estimation and ML concepts see \citealp{cowan98}).
Many efforts (e.g. \citealp{cash79,wachter79,baker84,gehrels86,nousek89,primini95,churazov96,jading96,
mighell99,hauschild01,bergmann02,arzner06}) have been devoted
to extend to the case of low count spectra the standard theories about curve fitting (best fit
parameters and confidence intervals estimation, goodness-of-fit test, etc.).
However no definitive solution has been found.


\section{The source plus background case}  \label{sec: real}

In this section we consider a more realistic situation by introducing a simplified instrumental background.
We model it with a power law (\verb|PEGPWRLW/b| within XSPEC), which is convolved with the RMF but not
multiplied by the ARF.
The power law slope, $\Gamma\mathrm{_{B}}$, is fixed to 0.25; the normalization, $N\mathrm{_{B}}$, is
calculated at the center of the energy band to minimize the correlation with $\Gamma\mathrm{_{B}}$.

There are two ways of analyzing spectra with background: we can subtract it using a spectrum from
blank field observations or we can model it.
When modeling the background, one can use the whole energy band (2.0-11.3~keV rather than 2.0-10.0~keV
as when using the subtraction) to increase the statistic.
Indeed, due to the high energy sharp decrease of the ARF of EPIC-MOS1, beyond 10~keV the thermal
component becomes negligible.
The background subtraction using the $\chi^2$ statistic is a widely used technique;
however, in the previous section (see $\S$\ref{sec: ideal}) we showed that, for low count spectra, the
Cash statistic is more suitable than the $\chi^2$ with reasonable grouping.
Since the Cash statistic requires the number of counts in each channel to be greater than zero,
the background has to be modeled \citep{cash79}.
We shall analyze simulated spectra in both ways and we shall compare the results.
Hereafter we call ``sub-$\chi^2$'' the standard analysis technique and ``mod-C'' the analysis using the
Cash statistic and the background modeling.

We proceed as for the source only case, considering as a guideline the Abell 1689 observation
mentioned in $\S$\ref{sec: ideal}. We produce a simulated spectrum choosing
realistic input values for an absorbed thermal (see $\S$\ref{sec: ideal}) plus background (see above)
model (\verb|WABS*MEKAL+PEGPWRLW/b| in XSPEC) and we produce $N\mathrm{_{meas}}$ different measurements
with a poissonian perturbation of the number of counts in each channel.
In the mod-C case, each spectrum is associated to the RMF and the ARF and is fitted between 2.0 and
11.3~keV with the \verb|WABS*MEKAL+PEGPWRLW/b| model.
The $N\mathrm{_H}$ and $\Gamma\mathrm{_{B}}$ are fixed to the input values, $z$ is allowed to vary between
0.186 and 0.214 ($\pm7\%$ of the input value), $kT$, $Z$, $N\mathrm{_{S}}$ and $N\mathrm{_{B}}$ are free.
We determine best fit values and one sigma uncertainties for all parameters.
Finally, we compute the weighted average of all $N\mathrm{_{meas}}$ values of each parameter
using one sigma uncertainties and we compare it with the input value.
In the sub-$\chi^2$ case, we simulate a background only spectrum with a long exposure time.
We consider a \verb|PEGPWRLW/b| model (slope and normalization are equal to those of the power law in the
source observation mentioned above) and we perturb the expected spectrum as explained above.
The adopted background spectrum is the same for all $N\mathrm{_{meas}}$ measurements and its exposure time
is 1\,000~ks.
We group the channels of each of the $N\mathrm{_{meas}}$ source spectra to have at least
25 counts per bin and we associate the background spectrum, the RMF and the ARF to the binned spectrum.
We fit the net spectrum with a thermal model only (\verb|WABS*MEKAL| in XSPEC) in the 2.0-10.0~keV band to
determine the best fit values, we compute the weighted averages and compare them with the input values.

\begin{figure*}
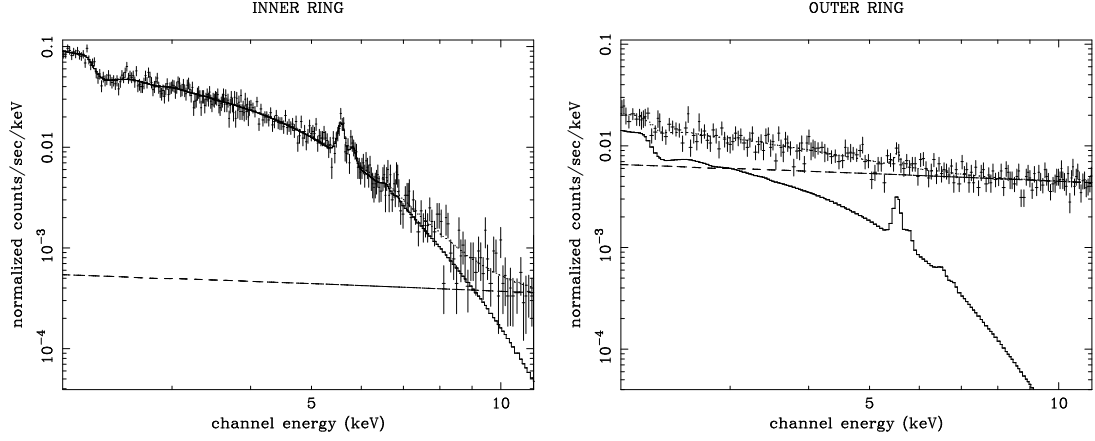

  \centering
  \begin{tabular}{cc}
    \resizebox{70mm}{!}{\includegraphics[angle=270]{inner_ring_spectrum.ps}} &
    \resizebox{70mm}{!}{\includegraphics[angle=270]{outer_ring_spectrum.ps}} \\
  \end{tabular}
  \caption{Simulated spectra accumulated in an inner ring (left panel, Fig.~1a) and in an outer
  ring (right panel, Fig.~1b). The solid line is the source contribution, the dashed line is the
  background and the dotted is the sum. In the outer ring, beyond 3~keV, background counts dominate
  over source counts. See text for further details and for model parameters.}
  \label{fig: real spec}
\end{figure*}

We consider two spatial regions: the ring between 1.0\arcmin~and 1.5\arcmin~centered on the cluster
emission peak, where the source dominates over the background (see Fig.~\ref{fig: real spec}a) and the
ring between 4.5\arcmin~and 6.0\arcmin, where the background dominates (see Fig.~\ref{fig: real spec}b).
Input values for the normalizations of both components are the best fit values measured in the two rings
of the Abell~1689 observation mentioned in $\S$\ref{sec: ideal}.
More specifically, in the inner ring $N\mathrm{_{S}} = 3.5\times10^{-3}$ and $N\mathrm{_{B}} = 1.5$
and in the outer ring $N\mathrm{_{S}} = 7.0\times10^{-4}$ and $N\mathrm{_{B}} = 17.5$ (XSPEC units).
For each ring we consider three input temperatures (namely 5, 7 and 9~keV) and two exposure times (10 and
100~ks).

\begin{table*}
  \caption{Comparison between the results obtained using the sub-$\chi^2$ and the mod-C
  data analysis techniques.}
  \label{tab: results sing}
  \centering
  \begin{tabular}{l r l c r c r}
    \hline \hline
     & & & \multicolumn{2}{c}{sub-$\chi^2$} & \multicolumn{2}{c}{mod-C} \\
    Ring & Exp. $\mathrm{^a}$ & $kT_0$ $\mathrm{^b}$ & $kT$ $\mathrm{^c}$ & $\Delta T/T_0$
    $\mathrm{^d}$ & $kT$ $\mathrm{^c}$ & $\Delta T/T_0$ $\mathrm{^d}$ \\
    \hline
    1.0\arcmin-1.5\arcmin & 100\,\, & 5.00 & 4.84$\pm$0.01 &  -3.2 \% & 4.96$\pm$0.01 &  -0.8\% \\
    1.0\arcmin-1.5\arcmin & 100\,\, & 7.00 & 6.78$\pm$0.02 &  -3.1 \% & 6.97$\pm$0.02 &  -0.4\% \\
    1.0\arcmin-1.5\arcmin & 100\,\, & 9.00 & 8.69$\pm$0.02 &  -3.4 \% & 8.97$\pm$0.03 &  -0.3\% \\
    \hline
    1.0\arcmin-1.5\arcmin & 10\,\,  & 5.00 & 4.81$\pm$0.03 &  -3.8 \% & 4.82$\pm$0.03 &  -3.6\% \\
    1.0\arcmin-1.5\arcmin & 10\,\,  & 7.00 & 6.78$\pm$0.05 &  -3.1 \% & 6.79$\pm$0.05 &  -3.0\% \\
    1.0\arcmin-1.5\arcmin & 10\,\,  & 9.00 & 8.68$\pm$0.11 &  -3.6 \% & 8.62$\pm$0.08 &  -4.2\% \\
    \hline
    4.5\arcmin-6.0\arcmin & 100\,\, & 5.00 & 3.95$\pm$0.01 & -21.0 \% & 4.71$\pm$0.02 &  -5.8\% \\
    4.5\arcmin-6.0\arcmin & 100\,\, & 7.00 & 5.24$\pm$0.02 & -25.1 \% & 6.45$\pm$0.03 &  -7.9\% \\
    4.5\arcmin-6.0\arcmin & 100\,\, & 9.00 & 6.43$\pm$0.02 & -28.6 \% & 8.09$\pm$0.04 & -10.1\% \\
    \hline
    4.5\arcmin-6.0\arcmin & 10\,\,  & 5.00 & 3.02$\pm$0.03 & -39.6 \% & 3.20$\pm$0.03 & -36.0\% \\
    4.5\arcmin-6.0\arcmin & 10\,\,  & 7.00 & 3.68$\pm$0.04 & -47.4 \% & 3.94$\pm$0.04 & -43.7\% \\
    4.5\arcmin-6.0\arcmin & 10\,\,  & 9.00 & 4.11$\pm$0.05 & -54.3 \% & 4.52$\pm$0.06 & -49.8\% \\
    \hline
  \end{tabular}
  \begin{list}{}{}
    \item[Notes:] $\mathrm{^a}$ exposure time in kiloseconds; $\mathrm{^b}$ input temperature in keV;
   $\mathrm{^c}$ measured temperature in keV; $\mathrm{^d}$ relative difference.
  \end{list}
\end{table*}

In Table~\ref{tab: results sing} we show the comparison between the two different data analysis
techniques described above (i.e. sub-$\chi^2$ and mod-C).
First we consider the inner ring, where the source dominates over the background.
The results are very similar to the case without background (see $\S$\ref{sec: ideal},
Table~\ref{tab: ideal1}).
For the 100~ks case mod-C returns the correct temperature and sub-$\chi^2$ slightly
underestimates it.
For shorter exposure times both techniques return a slightly biased value (bias $\approx$3\%).
No significant trend with the input temperature, $kT_0$, is found.
When considering the outer ring, where the background dominates, we find
different results: in all cases the true temperature is strongly underestimated.
There is a clear trend with the input temperature: the higher $kT_0$, the stronger the bias.
For long exposure times, mod-C (bias $\approx$10\%) works better than sub-$\chi^2$ (bias
$\approx$30\%).
Conversely, for short exposure times both techniques underestimate the true temperature by a
factor of about 2.
These results are qualitatively similar to those found for the source only case
(see Table~\ref{tab: ideal1}), but the bias is much stronger.

We have repeated the same analysis described above for a particular set of simulated spectra
(namely in the outer ring, with exposure time of 10~ks and $kT_0$~=~7~keV) modeling the source
with a bremsstrahlung rather than a \verb|MEKAL|.
The bremsstrahlung model is simpler and can be expressed as an analytic function of its two free
parameters (i.e. the temperature and the normalization).
Conversely, the \verb|MEKAL| model has two further parameters (metallicity and redshift) and its
complicated dependency on the parameters is not expressed in an analytic form: the expected values
are tabulated on a finite grid as a function of all parameters.
For this particular set of spectra, the bias for \verb|MEKAL| and bremsstrahlung models is the same.
This suggests that the bias is not related to the approximation of a finite grid of values.

Some insight on the origin of the bias can be gained by inspecting the probability density
function (p.d.f.) of the parameter of interest (in this case the temperature).
Here we deal with the Cash statistic, similar considerations apply to the $\chi^2$.
For each measurement, we define as $C\mathrm{_{min}}$ the absolute minimum value of the function $C(\alpha)$.
As in the previous section (see $\S$\ref{sec: ideal}) we minimize $C(\alpha)$ (see Eq.~\ref{eq: C})
to determine our best estimate, $\alpha\mathrm{_{best}}$, of the parameter set ($C\mathrm{_{min}} \equiv
C(\alpha\mathrm{_{best}})$).
Cash (\citeyear{cash79}) showed that the function $\Delta C$ (i.e. $C-C\mathrm{_{min}}$)
follows a $\chi^2$ distribution, therefore the confidence intervals can be generated in a standard way
(e.g. using the XSPEC command ERROR).
With the XSPEC command STEPPAR we produce the function $C$ for each free parameter
(here we consider only the temperature).
We calculate $\Delta C(T)$ and, since it is $\chi^2$ distributed, we can associate to each temperature,
$T\mathrm{_{X}}$, the probability that the true value is less or equal to $T\mathrm{_{X}}$,
i.e. the cumulative distribution function (c.d.f.) of the temperature, $P(T\mathrm{_{X}})$.
\begin{figure*}
  \centering
  \begin{tabular}{cc}
    \resizebox{70mm}{!}{\includegraphics{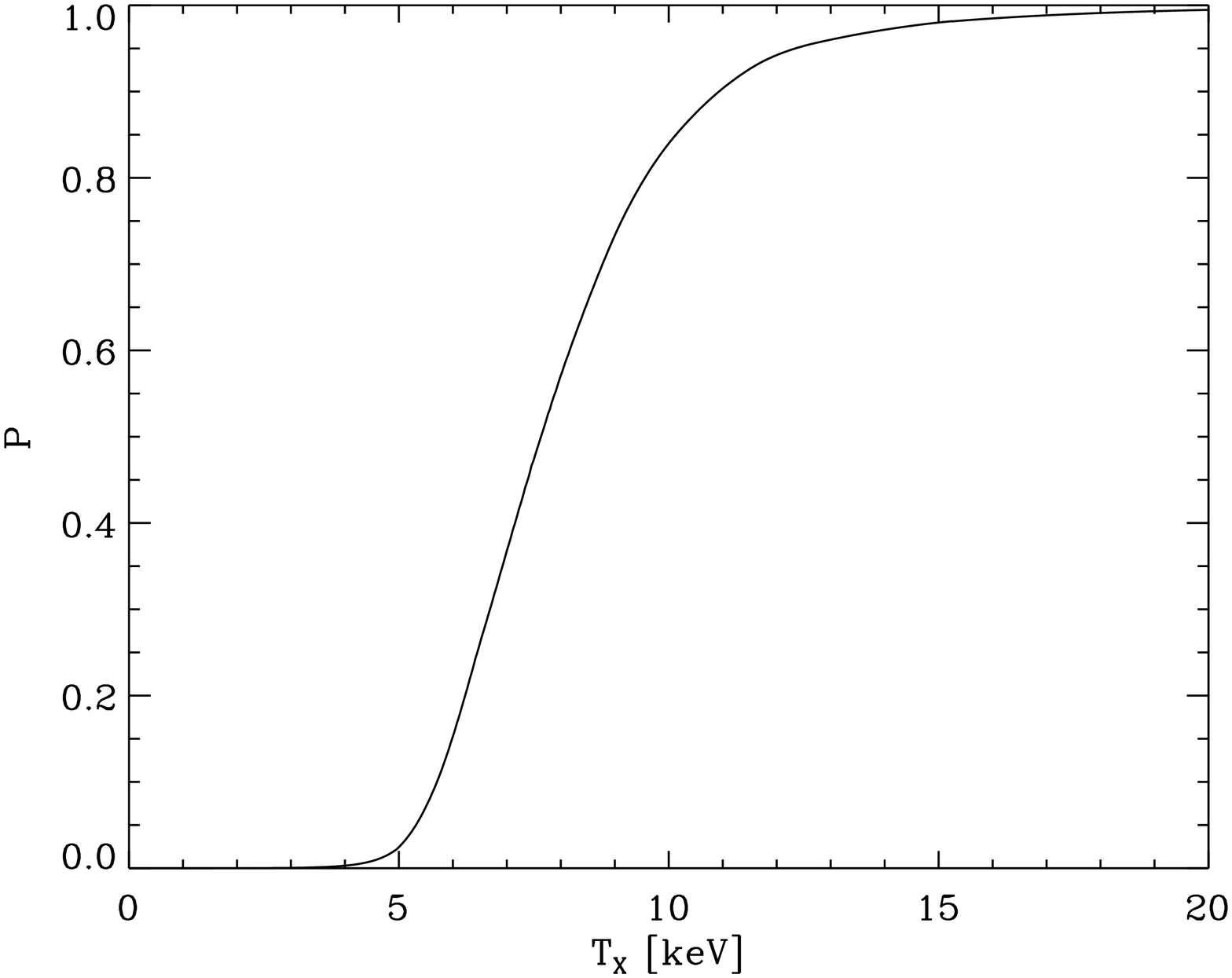}} &
    \resizebox{70mm}{!}{\includegraphics{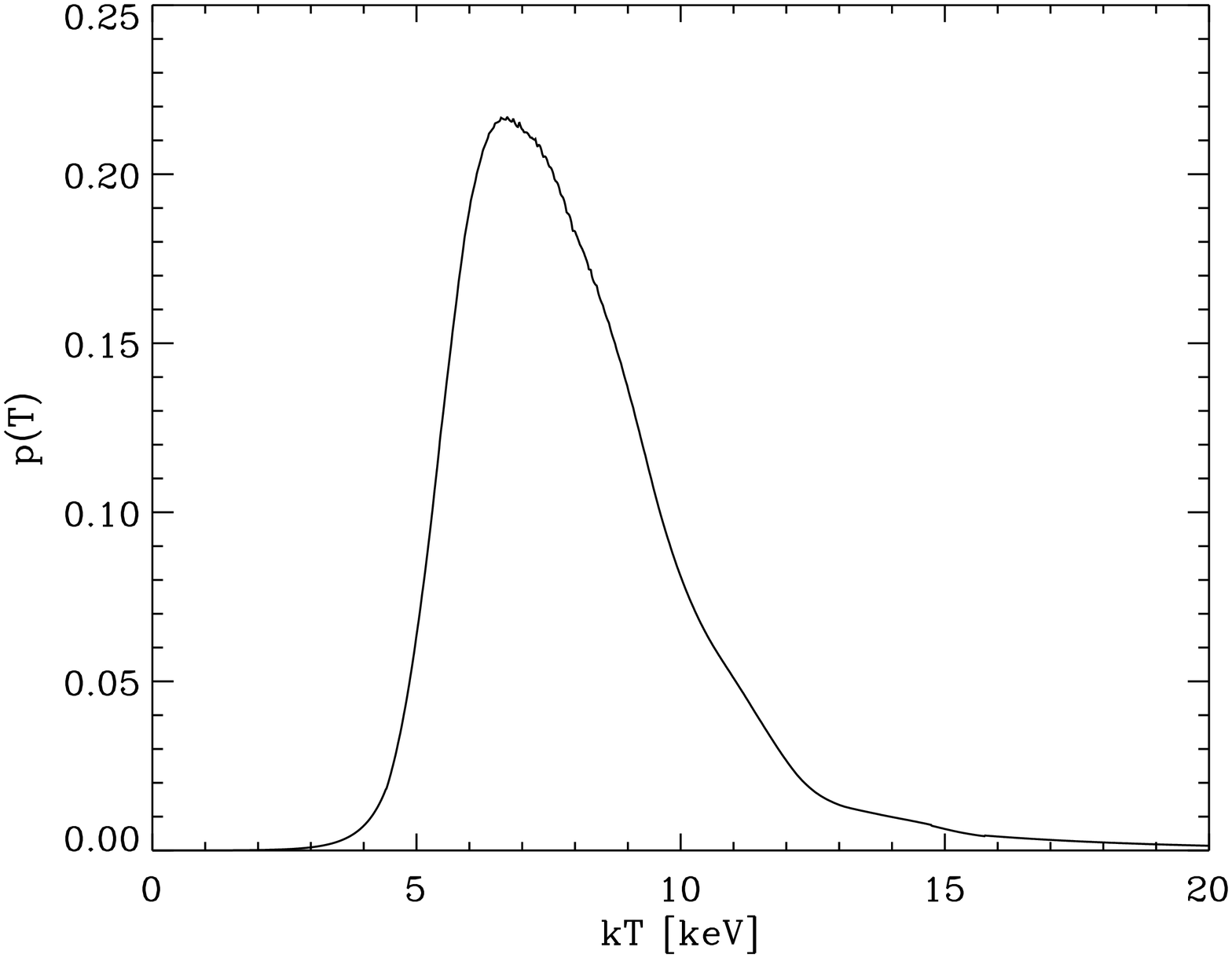}} \\
  \end{tabular}
  \caption{A cumulative distribution function (at left) and the associated probability density function
  (at right).}
  \label{fig: pdf pdd}
\end{figure*}
Given that $P(T\mathrm{_{X}}) = \int_0^{T\mathrm{_{X}}} p(T) \; \mathrm{d}T$, we can derive the
p.d.f. of the temperature, $p(T)$, for each single measurement (see Fig.~\ref{fig: pdf pdd}).
\begin{figure*}
  \centering
  \begin{tabular}{cc}
    \resizebox{70mm}{!}{\includegraphics{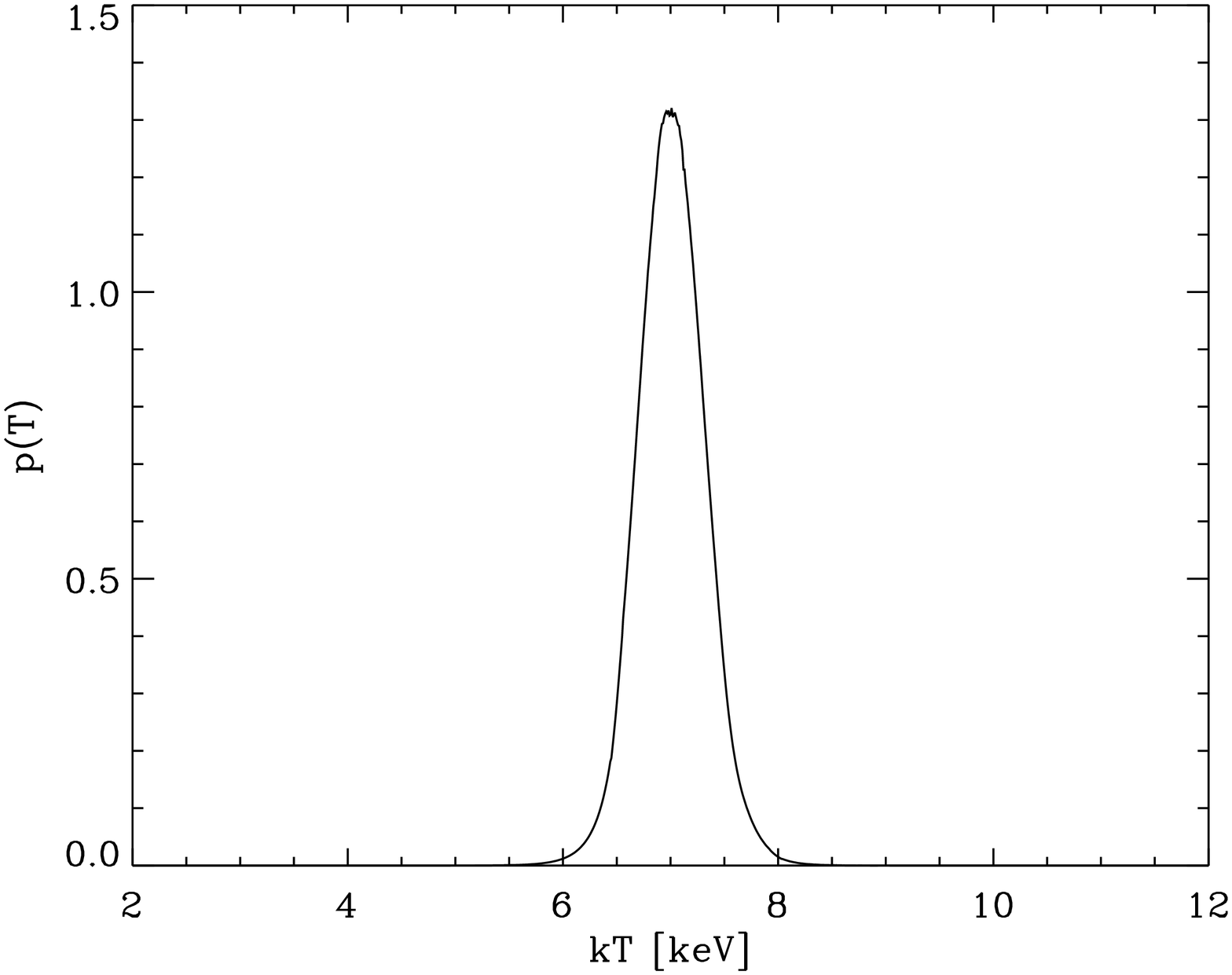}} &
    \resizebox{70mm}{!}{\includegraphics{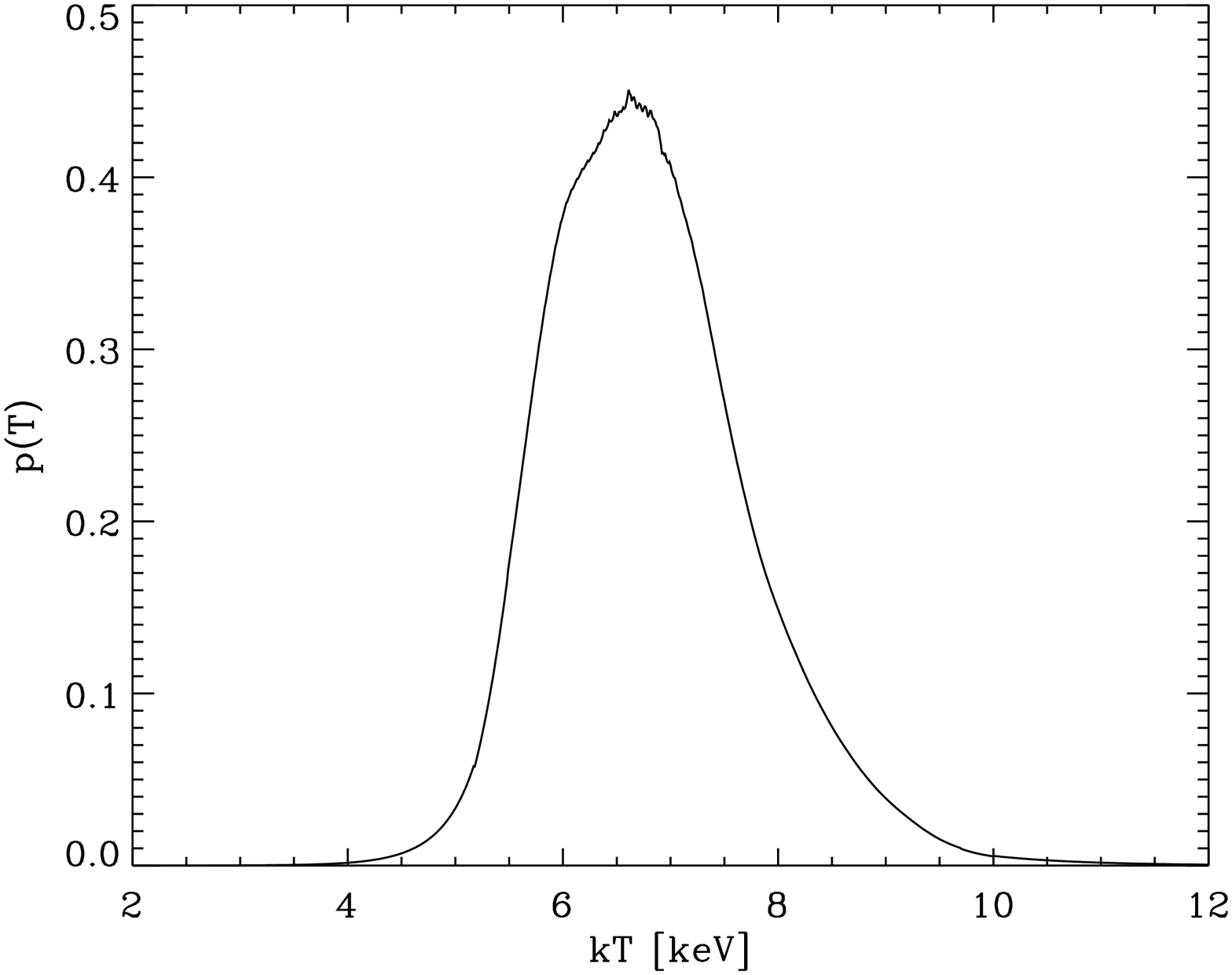}} \\
    \resizebox{70mm}{!}{\includegraphics{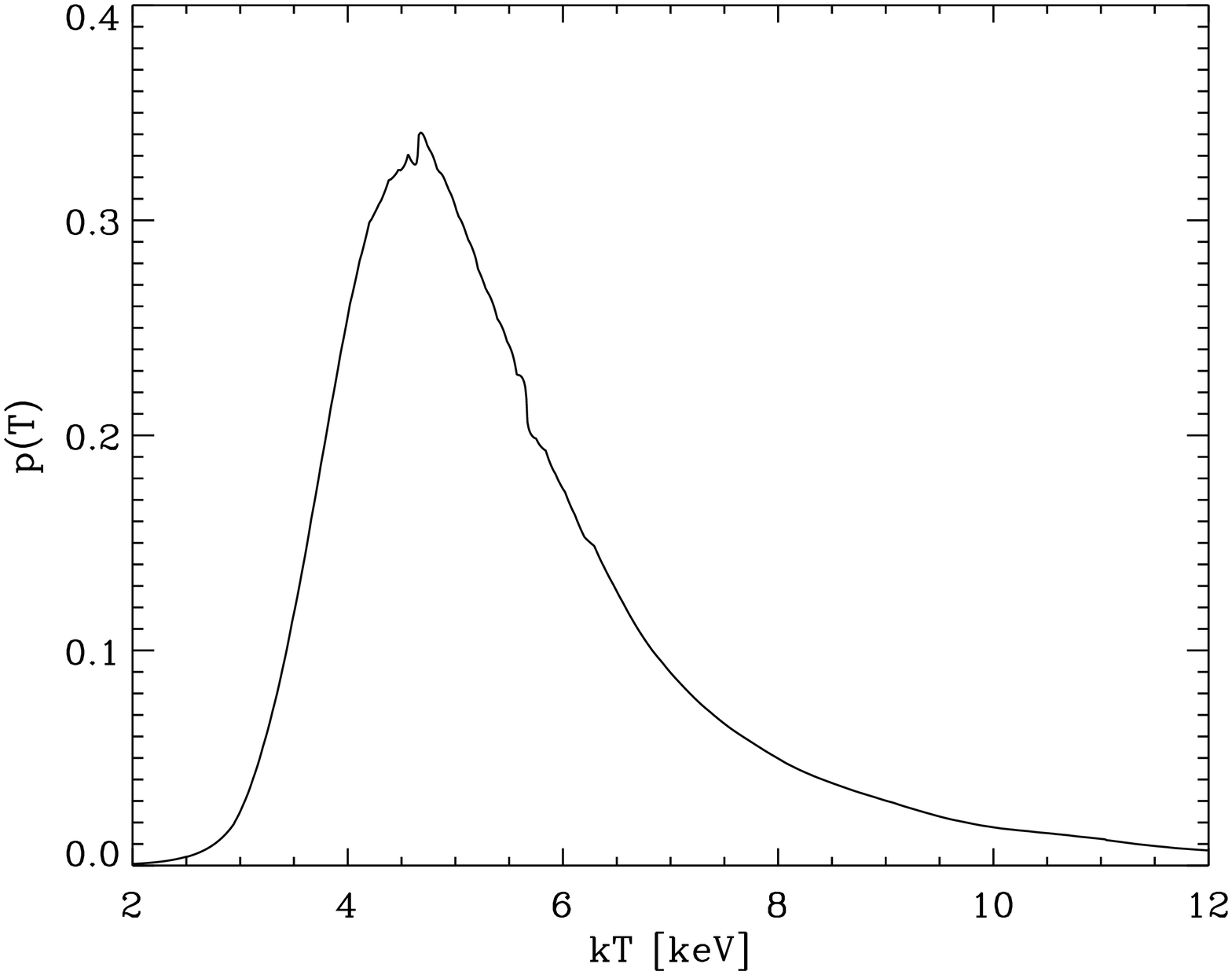}} &
    \resizebox{70mm}{!}{\includegraphics{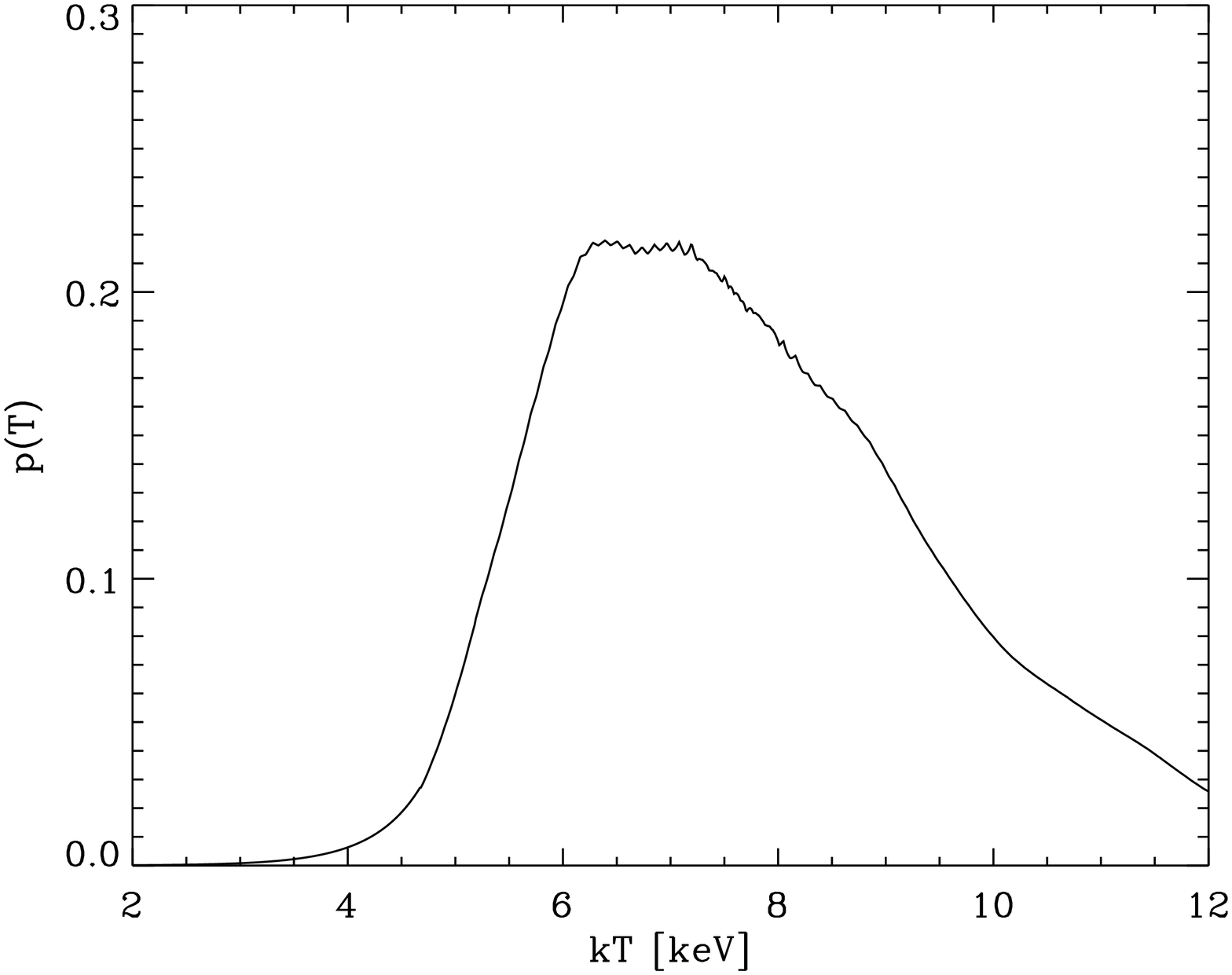}} \\
  \end{tabular}
  \caption{A comparison between representative $p(T)$ for single measurements extracted randomly
  in different conditions. The attention should be focused on the shapes of the p.d.f., rather
  than on temperature values. Top panels: the inner ring, where the source dominates over the
  background. Bottom panels: the outer ring, where the background is dominant. For left panels
  the exposure time is 100~ks, for right panels 10~ks. The input temperature is always 7~keV.
  Note that the scales in ordinate are different. Clearly the curves become less symmetric and less
  similar to gaussians, as the exposure time decreases and the background contribution increases.}
  \label{fig: prob distr}
\end{figure*}
In Fig.~\ref{fig: prob distr} we compare representative $p(T)$ for single measurements in
different conditions.
For each case the $p(T)$ is chosen randomly from the $N\mathrm{_{meas}}$ different measurements;
therefore, the attention should be focused on the shapes of the p.d.f., rather than on temperature values.
Clearly the curves become less symmetric and less similar to gaussians, as the exposure time
decreases and the background contribution increases.
The input temperature also plays a role: the higher is the temperature, the less symmetric is the curve.
Summarizing, the poorer the statistical quality of the data, the more asymmetric the $p(T)$,
the stronger the bias.

The way measurements are combined does not change the result.
We have experimented with two different methods: the weighted average of individual measurements
and the product of individual p.d.f..
The weighted average roughly approximates the $p(T)$ to a gaussian function and implies the
neglection of the contribution of high temperature tails.
A more appropriate way to join informations from different and independent measurements is to
multiply single p.d.f..
The best value for the parameter corresponds to the maximum of the joined p.d.f..
We multiply all $N\mathrm{_{meas}}$ p.d.f., computed as explained above,
and we still find a discrepancy between best fit and true values.
The bias is only slightly weaker than when computing a weighted average.
We have also tried computing the $p(T)$ in a different way, i.e. using the parametric bootstrap
technique \citep{press92}, which consists in creating and analyzing a large number of fake
datasets starting from model best fit values.
We obtain essentially the same results.

In Table~\ref{tab: results sing} we have showed that the strength of the bias mainly depends
on the total number of counts and on the background contribution.
A possibility to increase the total counts is to extend the band to lower energies.
We have explored it analyzing one of our set of simulated spectra (namely in the outer ring,
with exposure time of 10~ks and $kT_0$~=~7~keV) between 0.5 and 11.3~keV.
In this energy band the correlation between $kT$ and $N\mathrm{_{S}}$ is weaker and the
uncertainty on both parameters for a single measurement is smaller.
Using the mod-C technique, the bias in the broad band is smaller ($\approx$10\% vs. $\approx$40\%)
than in the narrow band; this suggests that also the parameter degeneracy could play an important
role when fitting in the 2.0-11.3 keV band.
However, in practice, it is not useful to enlarge the band to lower energies.
The main reasons are the imperfect calibration of EPIC instruments and the presence of the galactic
X-ray background (negligible beyond 2~keV), which introduce systematic effects that are hard to take
into account.
Moreover, broadband spectra are substantially more contaminated by emission from low temperature
components located on the same line of sight as the dominant component (see \citealp{mazzotta04,vikh05}).

In this section we have analyzed the realistic case of a thermal source with a background.
In such conditions a stronger bias is expected \citep{eadie71,bergmann02}.
As in the source only case (see $\S$\ref{sec: ideal}), we find that: 1) the $\chi^2$ and the Cash
estimators are strongly biased; 2) the Cash estimator is less biased than the $\chi^2$ one,
especially for long exposure times.
The strength of the bias depends mainly on two factors: the total number of counts and the
background contribution.


\section{Different attempts to correct the bias} \label{sec: correct bias}

Having established that neither sub-$\chi^2$ nor mod-C return acceptable results, we are faced with
two alternative ways to proceed.
A long term solution to the problem requires that an unbiased, or perhaps a less biased ML estimator,
be found and implemented within standard fitting packages (i.e. XSPEC).
Another, faster, solution involves correcting the bias a posteriori making use of extensive
montecarlo simulations.
In the following subsections we show our main results obtained exploring both approaches.

\subsection{Using different estimators} \label{subsec: estimators}

From the literature (e.g. \citealp{cowan98}) we know that even if $\hat{X}$ is an unbiased
estimator of $X$, $f(\hat{X})$ is not necessary an unbiased estimator of $f(X)$.
Reversing the argument one can argue that if $\hat{T}$ is a biased estimator of $T$, there may
exist a transformation, $f$, such that $f(\hat{T})$ is an unbiased (or at least less
biased) estimator of $f(T)$.
To test this idea we define in XSPEC (using the MDEFINE command) an analytic model similar to
bremsstrahlung, which we dub \verb|BREM2|:
\begin{equation}
S\mathrm{_T}(E) = N\mathrm{_S} \, E^{-4/3} \, T^{-1/2} \, \mathrm{exp}\left(-\frac{E}{T}\right) \; ,
\label{eq: brem2}
\end{equation}
where the energy, $E$, is expressed in keV; the normalization, $N\mathrm{_S}$, is chosen to reproduce
the same flux as a \verb|MEKAL| with no metals.

We simulate 3 sets of 3\,000 thermal plus background (\verb|BREM2+PEGPWRLW/b|) spectra with the
following input parameters: $N\mathrm{_S}$~=~$7.2 \times 10^{-4}$, $N\mathrm{_B}$~=~$17.5$
and $T$~=~7~keV.
These parameters correspond to those adopted in the case of the outer region (see $\S$\ref{sec: real}).
Each set has a different exposure time: 10, 20 and 100~ks.
We define 3 different estimators of the temperature: \\
$A$~=~$T^{-1} \;$,   \\
$B$~=~$T^{-1/2} \;$, \\
$C$~=~$T^{-1/4} \;$; \\
and their respective models: \\
$S\mathrm{_A}(E) = N\mathrm{_S} \, E^{-4/3} \, A^{1/2} \, \mathrm{exp}\left( -A \, E\right) \;$,  \\
$S\mathrm{_B}(E) = N\mathrm{_S} \, E^{-4/3} \, B \,       \mathrm{exp}\left(-B^2 \, E\right) \;$, \\
$S\mathrm{_C}(E) = N\mathrm{_S} \, E^{-4/3} \, C^{2} \,   \mathrm{exp}\left(-C^4 \, E\right) \;$. \\
For simplicity we have considered only power laws as different $f(T)$.
We fit each set of spectra with these models and measure the best estimate of $f(T)$.
We compute the weighted average of the 3\,000 $f(T)$ and calculate $T$ using the inverse function,
$f^{-1}$.
In Table~\ref{tab: estimators} we report the results of this analysis for different exposure times
and estimators.
\begin{table*}
  \caption{Comparison between the results obtained using different estimators of the temperature.
  The input temperature is 7 keV.}
  \label{tab: estimators}
  \centering
  \begin{tabular}{l c r c r c r}
    \hline \hline
     & \multicolumn{2}{c}{100 ks} & \multicolumn{2}{c}{20 ks} & \multicolumn{2}{c}{10 ks} \\
    Est. $\mathrm{^a}$ & $kT$ $\mathrm{^b}$ & $\Delta T/T_0$ $\mathrm{^c}$ & $kT$ $\mathrm{^b}$ &
    $\Delta T/T_0$ $\mathrm{^c}$ & $kT$ $\mathrm{^b}$ & $\Delta T/T_0$ $\mathrm{^c}$ \\
    \hline
    $T$        & 6.44$\pm$0.03 &  -8.0\% & 4.96$\pm$0.05 & -29.1\% & 4.04$\pm$0.07 & -42.3\% \\
    $T^{-1}$   & 6.96$\pm$0.03 &  -0.6\% & 7.46$\pm$0.08 &  +6.6\% & 8.83$\pm$0.14 & +26.1\% \\
    $T^{-1/2}$ & 6.88$\pm$0.03 &  -1.7\% & 6.59$\pm$0.06 &  -5.9\% & 6.36$\pm$0.09 &  -9.1\% \\
    $T^{-1/4}$ & 6.80$\pm$0.03 &  -2.9\% & 6.24$\pm$0.06 & -10.9\% & 5.72$\pm$0.08 & -18.3\% \\
    \hline
  \end{tabular}
  \begin{list}{}{}
    \item[Notes:] $\mathrm{^a}$ temperature estimator; $\mathrm{^b}$ measured temperature in keV;
    $\mathrm{^c}$ relative difference.
  \end{list}
\end{table*}
The choice of the estimator strongly affects the bias.
When using $T$ as estimator, we obtain results very similar to those obtained with a \verb|MEKAL|.
This is expected, because the model \verb|BREM2| is very similar to a bremsstrahlung
(see Eq.~\ref{eq: brem2}); note also that in $\S$\ref{sec: real} we showed that the bias is
roughly the same when considering a \verb|MEKAL| or a bremsstrahlung.
When considering the bias as a function of the power law index, we find a minimum corresponding to
$T^{-1/2}$, which is the best estimator among those considered.
For short exposure times (i.e. 10~ks) the use of $T^{-1/2}$ instead of $T$ reduces the bias by a
factor of 4.
We suggest that this could be related to the degree of complexity of the derivative of $S(E)$ with
respect to the estimator.
Note also that, when slightly increasing the statistic (e.g. when considering 20~ks of exposure
time), the bias associated to the $T^{-1/2}$ estimator is almost negligible if compared to
typical statistic uncertainties.
These results encourage the exploration of this approach (i.e. to consider different estimators)
in order to find a rigorously derived unbiased estimator of the temperature.

\subsection{Fitting with a log-normal function} \label{subsec: lognorm}

The shape of the $p(T)$ resembles the log-normal function, which is the p.d.f. of any random
variable whose logarithm is normally distributed. If $X$ is a random
variable with a normal distribution, then $x\equiv\exp(X)$ has a log-normal distribution.
The log-normal distribution has p.d.f.
\begin{equation}
f(x;\mu,\sigma) = \frac{1}{x \sigma \sqrt{2 \pi}} \; \mathrm{e}^{-(\ln x - \mu)^2/2\sigma^2}
\label{eq: lognorm_f}
\end{equation}
for $x \, > \, 0$, where $\mu$ and $\sigma$ are respectively the mean and the standard
deviation of the variable's logarithm.
The expected value is
\begin{equation}
\mathrm{E}(X) = \mathrm{e}^{\mu + \sigma^2/2}
\label{eq: lognorm_exp}
\end{equation}
and the variance is
\begin{equation}
\mathrm{var}(X) = (\mathrm{e}^{\sigma^2} - 1) \; \mathrm{e}^{2\mu + \sigma^2}.
\label{eq: lognorm_var}
\end{equation}

We fit each $p(T)$ with a log-normal function, $f(x;\mu,\sigma)$
(see Eq.~\ref{eq: lognorm_f}), and we calculate the best values of $\mu_i$ and $\sigma_i$.
We compute a weighted average of $\mu_i$ using $\sigma_i^{-2}$ as weights and calculate the
expected value (see Eq.~\ref{eq: lognorm_exp}) and the uncertainty, i.e. the variance
(see Eq.~\ref{eq: lognorm_var}) divided by the square root of the number of measurements.
The results are reported in Table~\ref{tab: results lognorm}.
In all cases, this method provides better results than a simple weighted average
(see Table~\ref{tab: results sing} for a comparison).
There is still a bias of a few percent, except for the case of the outer ring with 10~ks:
in this case the bias is greater than 10\%.
Thus, when the background contribution is small the log-normal distribution provides a good estimate,
while when the background is dominant the result is still biased, especially for few total counts,
but much less than when using the standard techniques.

\begin{table}
  \caption{Results obtained fitting $p(T)$ with a log-normal distribution.}
  \label{tab: results lognorm}
  \centering
  \begin{tabular}{l r l c r}
    \hline \hline
    Ring & Exp. $\mathrm{^a}$ & $kT_0$ $\mathrm{^b}$ & $kT$ $\mathrm{^c}$ &
    $\Delta T/T_0$ $\mathrm{^d}$ \\
    \hline
    1.0\arcmin-1.5\arcmin & 100\,\, & 5.00 & 4.96$\pm$0.01 &  -0.8 \% \\
    1.0\arcmin-1.5\arcmin & 100\,\, & 7.00 & 6.97$\pm$0.01 &  -0.4 \% \\
    1.0\arcmin-1.5\arcmin & 100\,\, & 9.00 & 8.97$\pm$0.01 &  -0.3 \% \\
    \hline
    1.0\arcmin-1.5\arcmin & 10\,\,  & 5.00 & 4.88$\pm$0.02 &  -2.4 \% \\
    1.0\arcmin-1.5\arcmin & 10\,\,  & 7.00 & 6.90$\pm$0.05 &  -1.4 \% \\
    1.0\arcmin-1.5\arcmin & 10\,\,  & 9.00 & 8.81$\pm$0.13 &  -2.1 \% \\
    \hline
    4.5\arcmin-6.0\arcmin & 100\,\, & 5.00 & 4.90$\pm$0.02 &  -2.0 \% \\
    4.5\arcmin-6.0\arcmin & 100\,\, & 7.00 & 6.77$\pm$0.04 &  -3.3 \% \\
    4.5\arcmin-6.0\arcmin & 100\,\, & 9.00 & 8.51$\pm$0.09 &  -5.4 \% \\
    \hline
    4.5\arcmin-6.0\arcmin & 10\,\,  & 5.00 & 4.68$\pm$0.13 &  -6.4 \% \\
    4.5\arcmin-6.0\arcmin & 10\,\,  & 7.00 & 5.90$\pm$0.24 & -15.7 \% \\
    4.5\arcmin-6.0\arcmin & 10\,\,  & 9.00 & 7.67$\pm$0.51 & -14.8 \% \\
    \hline
  \end{tabular}
  \begin{list}{}{}
    \item[Notes:] $\mathrm{^a}$ exposure time in kiloseconds; $\mathrm{^b}$ input temperature in keV;
   $\mathrm{^c}$~measured temperature in keV; $\mathrm{^d}$ relative difference.
  \end{list}
\end{table}

\subsection{A semi-empirical method: summing three distributions} \label{subsec: triplet}

\begin{figure}
  \centering
  \resizebox{80mm}{!}{\includegraphics{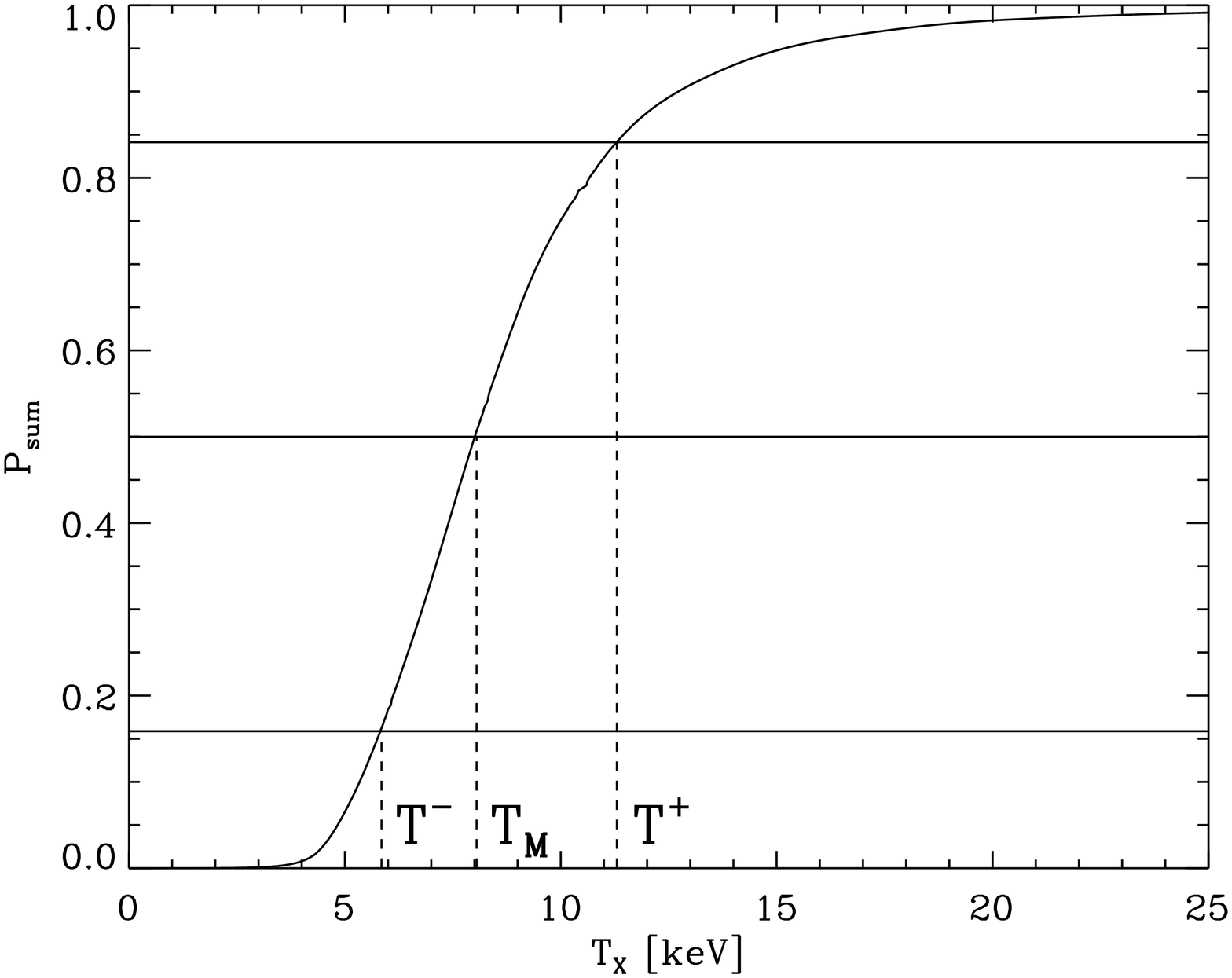}}
  \caption{Visual representation of the definition of $T^-$, $T\mathrm{_M}$ and $T^+$ from
  a joined cumulative distribution function.}
  \label{fig: how}
\end{figure}

The three EPIC instruments (MOS1, MOS2 and pn) on board \emph{XMM-Newton} provide three
simultaneous and independent measurements of the same target; therefore, when dealing with EPIC data,
one has the necessity of correctly combining these three measurements.
A weighted average is the simplest procedure, however in $\S$\ref{sec: real} we showed that it leads
to biased results.
In $\S$\ref{sec: real} we also showed that the strength of the bias is related to the shape of the
p.d.f. and in $\S$\ref{subsec: lognorm} we showed that a fit with a log-normal function does not
return sufficiently accurate results; in this section we try to proceed in a different way,
emphasizing the contribution of p.d.f. tails at high temperatures.
We derive $N\mathrm{_{meas}}$ measurements of the temperature with their corresponding p.d.f., as in the
mod-C case described in $\S$\ref{sec: real}.
We divide the $N\mathrm{_{meas}}$ measurements in groups of three and, for each group, we consider the
three p.d.f., $p\mathrm{_i}(T)$ (i=1,2,3), and combine them in a non-standard way: we calculate the sum,
rather than the product, of the single p.d.f..
In practice, it is equivalent, but more useful, to sum directly the c.d.f., $P\mathrm{_i}(T\mathrm{_X})$.
The sum is renormalized dividing it by 3.
We define $P\mathrm{_{sum}}(T\mathrm{_X})$ as follows:
\begin{equation}
P\mathrm{_{sum}}(T\mathrm{_X}) = \frac{1}{3} \sum_{i=1}^3 P\mathrm{_i}(T\mathrm{_X}) \; .
\end{equation}
This is a sort of joined c.d.f. of three measurements and the associated p.d.f. is usually
more symmetric than the single $p\mathrm{_i}(T)$.
We define as $T^-$, $T\mathrm{_M}$ and $T^+$ the temperatures which correspond to a probability, 
$P\mathrm{_{sum}}(T\mathrm{_X})$, of 0.1587, 0.5000 and 0.8413 respectively (see Fig.~\ref{fig: how}).
We consider $T\mathrm{_M}$ as the best estimate for the three joined measurements,
$\mathrm{d}T^- \equiv (T\mathrm{_M}-T^-)/\sqrt{3}$ as the lower uncertainty and
$\mathrm{d}T^+ \equiv (T^+-T\mathrm{_M})/\sqrt{3}$ as the upper uncertainty.
Thus we have $N\mathrm{_{meas}}/3$ ``triplet'' measurements: $T\mathrm{_M}^{+\mathrm{d}T^+}_{-\mathrm{d}T^-}$.
We compute the weighted average of the $N\mathrm{_{meas}}/3$ ``triplets'' and
in Table~\ref{tab: results triplets} we compare the results with the input values.
\begin{table}
  \caption{Results obtained with the semi-empirical ``triplet'' method.}
  \label{tab: results triplets}
  \centering
  \begin{tabular}{l r l c r}
    \hline \hline
    Ring & Exp. $\mathrm{^a}$ & $kT_0$ $\mathrm{^b}$ & $kT$ $\mathrm{^c}$ & $\Delta T/T_0$
    $\mathrm{^d}$ \\
    \hline
    1.0\arcmin-1.5\arcmin & 100\,\, & 5.00 & 4.97$\pm$0.01 & -0.6 \% \\
    1.0\arcmin-1.5\arcmin & 100\,\, & 7.00 & 6.97$\pm$0.02 & -0.4 \% \\
    1.0\arcmin-1.5\arcmin & 100\,\, & 9.00 & 8.98$\pm$0.03 & -0.2 \% \\
    \hline
    1.0\arcmin-1.5\arcmin & 10\,\,  & 5.00 & 4.90$\pm$0.04 & -2.0 \% \\
    1.0\arcmin-1.5\arcmin & 10\,\,  & 7.00 & 6.94$\pm$0.07 & -0.9 \% \\
    1.0\arcmin-1.5\arcmin & 10\,\,  & 9.00 & 8.98$\pm$0.11 & -0.2 \% \\
    \hline
    4.5\arcmin-6.0\arcmin & 100\,\, & 5.00 & 5.00$\pm$0.02 & -0.0 \% \\
    4.5\arcmin-6.0\arcmin & 100\,\, & 7.00 & 6.90$\pm$0.04 & -1.4 \% \\
    4.5\arcmin-6.0\arcmin & 100\,\, & 9.00 & 8.91$\pm$0.05 & -1.0 \% \\
    \hline
    4.5\arcmin-6.0\arcmin & 10\,\,  & 5.00 & 5.04$\pm$0.06 & +0.8 \% \\
    4.5\arcmin-6.0\arcmin & 10\,\,  & 7.00 & 6.97$\pm$0.09 & -0.4 \% \\
    4.5\arcmin-6.0\arcmin & 10\,\,  & 9.00 & 8.96$\pm$0.13 & -0.4 \% \\
    \hline
  \end{tabular}
  \begin{list}{}{}
    \item[Notes:] $\mathrm{^a}$ exposure time in kiloseconds; $\mathrm{^b}$ input temperature in keV;
   $\mathrm{^c}$~measured temperature in keV; $\mathrm{^d}$ relative difference.
  \end{list}
\end{table}
In almost all cases this semi-empirical method (hereafter ``triplet'' method) provides excellent
results: the discrepancy is lower than 2\% and often comparable with the statistical uncertainty.

We have tried joining different numbers of measurements together; simulations show that,
when considering two measurements at a time, we find the temperature to be underestimated, when
considering five measurements, we obtain substantially correct results, as when using the ``triplets''.
This suggest that the effectiveness of our a posteriori correction depends on the number of
measurements we combine.
We propose to use the ``triplets'' because three is the minimum number of measurements for which we
obtain unbiased temperature estimates and because this is a natural choice when analyzing
EPIC data.

We want to stress that this technique is not rigorously derived from statistics principles, but we
have showed that it is the only method that returns the expected temperature under very
different situations (e.g. different background contributions and exposure times).
This could be related to the fact that joined p.d.f. are usually more symmetric than the single ones.


\section{Summary} \label{sec: conclusions}
It is well known from the literature (e.g \citealp{eadie71} or \citealp{cowan98}) that ML estimators
are generally biased and that they are gaussian and unbiased only in the asymptotic limit.
In this paper we test the effects of statistical fluctuations in determining the temperature from
a thermal spectrum.
In particular, we explore a range of conditions for which ML estimators reveal their intrinsic bias.

In the source only case we show that:
\begin{enumerate}
   \item the estimators of the temperature based on the Cash and the $\chi^2$ statistics are
         biased for short exposure times, i.e. for few counts;
   \item the Cash statistic works better than the $\chi^2$, as pointed out by \citet{cash79};
   \item the $\chi^2$ statistic works as well as the Cash, when strongly increasing
         channel grouping.
\end{enumerate}
In the source plus background case we show that:
\begin{enumerate}
   \item the standard analysis techniques (sub-$\chi^2$ and mod-C) return heavily biased results;
   \item the strength of the bias depends mainly on the total number of counts and on the
         background contribution;
   \item the use of different estimators of the temperature, in particular $T^{-1/2}$, strongly
         reduces the bias.
\end{enumerate}
The last point is encouraging in order to find a long term solution of the problem (i.e. a
rigorously derived unbiased estimator).
As an alternative and immediate solution we propose the so-called ``triplet'' method, which
makes use of a correction of the bias a posteriori, working on the probability distribution
functions.
This semi-empirical recipe returns the correct result under very different situations,
even though it is not rigorously derived from statistics principles.

We point out that our results can have strong implications for the measurement of the temperature
from spectra accumulated from low surface brightness regions (e.g. the outer regions of galaxy
clusters) with current experiments, i.e. \emph{XMM-Newton} and \emph{Chandra} (see for
example \citealp{piffa05,vikh05,pratt07}).
Indeed the bias seems to be related to the statistical quality of the data, which typically depends
on the distance of a given region from the core.
For this reason we might expect a net effect on radial temperature profiles of galaxy clusters.
In a forthcoming paper \citep{leccardi07} we will analyze a sample of clusters to determine
the mean temperature profile using our data analysis technique.


\begin{acknowledgements}
We would like to thank an anonymous referee for helpful comments.
We thank A.~Finoguenov for useful discussions.
\end{acknowledgements}


\bibliographystyle{aa}
\bibliography{paper1}

\begin{thebibliography}{20}
\expandafter\ifx\csname natexlab\endcsname\relax\def\natexlab#1{#1}\fi

\bibitem[{{Arzner} {et~al.}(2006){Arzner}, {G{\"u}del}, {Briggs}, {Telleschi},
  {Schmidt}, {Audard}, {Scelsi}, \& {Franciosini}}]{arzner06}
{Arzner}, K., {G{\"u}del}, M., {Briggs}, K., {et~al.} 2006, astro-ph/0609193

\bibitem[{{Baker} \& {Cousins}(1984)}]{baker84}
{Baker}, S. \& {Cousins}, R.~D. 1984, Nucl. instr. and meth. A, 221, 437

\bibitem[{{Bergmann} \& {Riisager}(2002)}]{bergmann02}
{Bergmann}, U.~C. \& {Riisager}, K. 2002, Nucl. instr. and meth. A, 489, 444

\bibitem[{Cash(1979)}]{cash79}
Cash, W. 1979, \apj, 228, 939

\bibitem[{Churazov {et~al.}(1996)Churazov, Gilfanov, Forman, \&
  Jones}]{churazov96}
Churazov, E., Gilfanov, M., Forman, W., \& Jones, C. 1996, \apj, 471, 673

\bibitem[{{Cowan}(1998)}]{cowan98}
{Cowan}, G. 1998, {Statistical data analysis} (Oxford Science Publications)

\bibitem[{{Eadie} {et~al.}(1971){Eadie}, {Drijard}, {James}, {Roos}, \&
  {Sadoulet}}]{eadie71}
{Eadie}, W.~T., {Drijard}, D., {James}, F.~E., {Roos}, M., \& {Sadoulet}, B.
  1971, {Statistical Methods in Experimental Physics} (Amsterdam: North-Holland
  Publishers)

\bibitem[{{Gehrels}(1986)}]{gehrels86}
{Gehrels}, N. 1986, \apj, 303, 336

\bibitem[{{Hauschild} \& {Jentschel}(2001)}]{hauschild01}
{Hauschild}, T. \& {Jentschel}, M. 2001, Nucl. instr. and meth. A, 457, 384

\bibitem[{{Jading} \& {Riisager}(1996)}]{jading96}
{Jading}, Y. \& {Riisager}, K. 1996, Nucl. instr. and meth. A, 372, 289

\bibitem[{{Kearns} {et~al.}(1995){Kearns}, {Primini}, \&
  {Alexander}}]{primini95}
{Kearns}, K., {Primini}, F., \& {Alexander}, D. 1995, in ASP Conf. Ser. 77:
  Astronomical Data Analysis Software and Systems IV, ed. R.~A. {Shaw}, H.~E.
  {Payne}, \& J.~J.~E. {Hayes}, 331--+

\bibitem[{{Leccardi} \& {Molendi}(2007)}]{leccardi07}
{Leccardi}, A. \& {Molendi}, S. 2007, in preparation

\bibitem[{{Mazzotta} {et~al.}(2004){Mazzotta}, {Rasia}, {Moscardini}, \&
  {Tormen}}]{mazzotta04}
{Mazzotta}, P., {Rasia}, E., {Moscardini}, L., \& {Tormen}, G. 2004, \mnras,
  354, 10

\bibitem[{{Mighell}(1999)}]{mighell99}
{Mighell}, K.~J. 1999, \apj, 518, 380

\bibitem[{{Nousek} \& {Shue}(1989)}]{nousek89}
{Nousek}, J.~A. \& {Shue}, D.~R. 1989, \apj, 342, 1207

\bibitem[{Piffaretti {et~al.}(2005)Piffaretti, Jetzer, Kaastra, \&
  Tamura}]{piffa05}
Piffaretti, R., Jetzer, P., Kaastra, J.~S., \& Tamura, T. 2005, \aap, 433, 101

\bibitem[{Pratt {et~al.}(2007)Pratt, B{\"o}hringer, Croston, Arnaud, Borgani,
  Finoguenov, \& Temple}]{pratt07}
Pratt, G.~W., B{\"o}hringer, H., Croston, J.~H., {et~al.} 2007, \aap, 461, 71

\bibitem[{Press {et~al.}(1992)Press, Flannery, Teukolsky, \&
  Vetterling}]{press92}
Press, W.~H., Flannery, B.~P., Teukolsky, S.~A., \& Vetterling, W.~T. 1992,
  {Numerical recipes} (Cambridge University Press)

\bibitem[{Vikhlinin {et~al.}(2005)Vikhlinin, Markevitch, Murray, Jones, Forman,
  \& Van~Speybroeck}]{vikh05}
Vikhlinin, A., Markevitch, M., Murray, S.~S., {et~al.} 2005, \apj, 628, 655

\bibitem[{{Wachter} {et~al.}(1979){Wachter}, {Leach}, \& {Kellogg}}]{wachter79}
{Wachter}, K., {Leach}, R., \& {Kellogg}, E. 1979, \apj, 230, 274

\end{thebibliography}

\end{document}